\documentstyle[aps,preprint]{revtex}
\begin{document}
\draft
\title{Frequency Locking in Spatially Extended Systems}
\author{Hwa-Kyun Park\footnote{e-mail: childend@front.kaist.ac.kr \\
 TEL:+82-42-869-2563 \hspace{.5cm} FAX: +82-42-869-2510}} 
\address{Department of Physics,Korea Advanced Institute of Science and
Technology,Taejon 305-701,Korea} 
\date{25 April 2000}

\maketitle

\begin{abstract}

A variant of the complex Ginzburg-Landau equation is used to 
investigate the frequency locking phenomena in spatially 
extended systems.
With appropriate parameter values, a variety of
frequency-locked patterns including flats, $\pi$ fronts, labyrinths
and $2\pi/3$ fronts emerge.
We show that in spatially extended systems, frequency locking
can be enhanced or suppressed by diffusive coupling.
Novel patterns such as chaotically bursting domains and 
target patterns are also observed during the transition to locking. 
\end{abstract}


Frequency locking of nonlinear oscillators is a well known phenomenon, 
and has been studied extensively\cite{ott,mann};
however, frequency locking in spatially extended systems is not well
understood. 
Recently, frequency locking has been observed experimentally 
in the light sensitive Belousov-Zhabotinsky reaction\cite{pos97,lps99}.
The system is oscillatory, and shows spiral patterns without 
external forcing. When periodically illuminated with light pulses,
however, it exhibits frequency-locked patterns such as flats, 
$\pi$ fronts, labyrinths and $2\pi/3$ fronts, depending on the
forcing frequency.
It was also reported that these resonances have the structure similar 
to Arnol'd tongues\cite{lbm00}.

For the theoretical investigation of frequency locking in spatially
extended systems, we consider the forced complex Ginzburg-Landau
equation \cite{ce92,zv95,cfs94,ta97,cl90,eh98,eh99,lha00}

\begin{equation}\label{fcgl}
\partial_t \psi= (1 - i\nu) \psi + (-1 +i b) |\psi|^2 \psi
+ (1+id)\nabla^2 \psi + c \overline{\psi}^{n-1},
 \end{equation}
where $\psi(\vec{r},t)$ is a complex order parameter, and 
$\nu, b, d, c$ are real.

A theoretical study on the Eq. (\ref{fcgl})
showed that resonant forcing can exhibit rich patterns as varying 
the external frequency\cite{ce92}.
The same equation has been studied for parametric resonance cases
in the system such as Faraday waves \cite{zv95,cfs94} and vertically 
oscillated granular layers \cite{ta97}. 
However these systems are turbulent\cite{ce92} or 
stationary\cite{zv95,cfs94,ta97} without forcing.
For the oscillatory cases, the studies have been focused 
on the front dynamics in 2:1\cite{cl90} and $2n:1$
\cite{eh98,eh99,lha00} resonance cases.  
More works are required to understand the frequency locking 
phenomena in spatially extended system.

In this letter, we analyze the Arnol'd tongue structures 
of spatiotemporal patterns for 1:1,2:1 and 3:1 resonance cases.
Especially, we describe how diffusive couplings affect the
frequency locking.

Firstly, we will explain the model and parameter values.
Eq. (\ref{fcgl}) with $c=0,\nu=0$ is a well-known complex 
Ginzburg-Landau equation, which describes oscillatory media
and shows stably rotating spiral patterns when 
$1-bd>0$\cite{kura,mik,cm96}:

\begin{equation}\label{cgl}
\partial_t \psi= \psi + (1+id)\nabla^2 \psi + (-1 +i b) |\psi|^2 \psi
.
\end{equation}
In this case, the model is considered as the amplitude
equation of chemical oscillations $h_0=\psi e^{iwt} + c.c.$, where $w$
is the frequency of a spiral rotation in real experiments.

Introducing a local amplitude $A(\vec{r},t)$ and a local phase
$\phi(\vec{r},t)$ by $\psi (\vec{r},t)=A(\vec{r},t)
\exp(i\phi (\vec{r},t))$, Eq. (\ref{cgl}) reduces to the following
phase equation\cite{kura,mik}:

\begin{equation}\label{phase0}
\partial_t \phi= b+(1-bd)\nabla^2 \phi -(b+d)(\nabla \phi)^2
.
\end{equation}
It has a oscillatory solution
$\phi=(b-(b+d)k^2) t+  \vec{k} \cdot \vec{r} +\phi_0$.  
Diffusion coefficient $1-bd$ must be positive for stable phase
dynamics.
Therefore, two constraints on parameter values, $1-bd>0$ and $b+d < 0$, 
are required from the experimental facts that without forcing
there are stably rotating spiral patterns and its frequency is 
higher than that of a homogeneous system\cite{pos97}.
Specifically, we choose $b=-0.5,d=-1.4$.

Now, consider when the spirals are forced with external
frequency $w_e = n(w+\nu)$, 
where $n$ is an irreducible integer fraction
and $\nu$ is a small deviation from a rational 
multiple of the natural frequency $w$. 
In this case, $\psi$ is taken as the complex amplitude 
of chemical oscillation $h=\psi e^{i(w+\nu)t} + c.c.$,
so that stationary $\psi$ may indicate the entrainment of system
to the external frequency $w_e$. Correspondence between $h$ and 
$h_0=\psi e^{iwt} + c.c.$ when the forcing is absent gives 
the term $-i\nu \psi$ in Eq. (\ref{fcgl}).
Considering the invariance of the system under a translation 
$t \rightarrow t+ \frac{2\pi}{w_e}$ yields the resonant forcing 
term $c \overline{\psi}^{n-1}$ in Eq. (\ref{fcgl})

Next, we will obtain the Arnol'd tongue structure of 
Eq. (\ref{fcgl}) and study the characteristic of resonant
patterns.  
Eq. (\ref{fcgl}) contains two control parameters $\nu$
(measuring frequency mismatch) and $c$ (forcing amplitude).
In this parameter space, we seek the stationary solutions of
Eq. (\ref{fcgl}), which means frequency locking.
Without diffusion the linear stability analysis of 
the stationary state yields the Arnol'd tongue of 
a single oscillator (given as solid lines in Fig.\ref{fig1}).

There are $n$ equivalent stable stationary solutions
for $n:1$ resonances. These solutions have the same amplitudes,
but are phase-shifted with respect to each other. 
In spatially extended systems, different oscillators may 
lock to $n$  different states to make $2\pi /n$ front
solutions. 

To obtain the resonance regions of spatiotemporal patterns,
we integrate Eq. (\ref{fcgl}) numerically.
System size is 100 by 100 with periodic boundary conditions.
The initial condition is a set of Gaussian distributed random numbers
of mean zero, and variance 0.25.
The numerical results are summarized in Fig.\ref{fig1}; 
the typical patterns are presented in Fig.\ref{fig2}.

When the frequency locking does not occur, basic patterns are
spirals.
At 1:1 resonance, resonant patterns are homogeneous flats
(Fig.\ref{fig2}(a)). At 2:1 resonance, 
there are two resonant patterns: $\pi$ fronts
(Fig.\ref{fig2}(b)) and
labyrinthine patterns (Fig.\ref{fig2}(d)). 
Fig.\ref{fig1}(b) shows that labyrinthine patterns arise 
at higher frequencies than $\pi$ fronts. 
Under some conditions $\pi$ fronts move until 
only one domain remains. These front dynamics and 
the nonequilibrium Ising-Bloch transition(NIB) have 
been studied by previous researchers\cite{lbm00,cl90,pgs99}.  
At 3:1 resonance, we can see the $2\pi /3$ fronts as in 
Fig.\ref{fig2}(c).
But the time series at one typical point show that these fronts
are in fact not stationary, but move slowly(Fig.\ref{fig3}(a)). 
The periodic nature of time series indicates the motion of the
fronts. The flat region of time series corresponds to the time
interval during which each domain passes. 
Sometimes moving fronts merge and eventually collapse into 
flat patterns.
Ref .\cite{cl90} shows that chiral interfaces move in
nonequilibrium systems. In this respect, observed $2\pi /3$ 
front motion can be explained from the nonzero chirality of the 
interfaces (Fig.\ref{fig3}(b)).

Next, we discuss how diffusion affects frequency locking
and Arnol'd tongue structures (See Fig.\ref{fig1}).
The regions indicated by 1F, 2F, 3F and LABY, which corresponds
to flat, $\pi$ fronts, $2\pi/3$ fronts and labyrinths, respectively,
are frequency locked. 
The Arnol'd tongue structure in Eq. (\ref{fcgl}) 
almost coincides with that of a single oscillator,
but some deviations are found.
For example, the tip of Arnol'd tongue is smoothed 
so that locking begins at a nonzero forcing amplitude.
It is rather surprising that frequency-locked labyrinths
arise at parameter values at which a single oscillator 
without coupling would not be locked.
It means that diffusion induces frequency locking.
One can also see that at lower $\nu$, the locking is somewhat 
suppressed. (Note that the regions indicated by BD are not 
frequency-locked.)

To understand the enhancement and suppression of locking,
we consider the phase dynamics of Eq. (\ref{fcgl}).
Substituting $\psi(\vec{r},t)=A(\vec{r},t) \exp(i\phi (\vec{r},t))$
into Eq. (\ref{fcgl}) and assuming that the amplitude is
slaved to the phase, we get the following phase equation 
for $n=2$:

\begin{equation}\label{phase}
\partial_t \phi = (b-\nu)+(1-bd)\nabla^2 \phi -(b+d)(\nabla
\phi)^2 + c\sqrt{1+b^2} \cos(2\phi + \delta) 
,
\end{equation}
where $\tan(\delta)=1/b$.

Eq. (\ref{phase}) has an oscillatory solution 
$\phi=(b-\nu-(b+d)k^2) t+ \vec{k} \cdot \vec{r} +\phi_0$ for $c=0$.
The stationary solution for nonzero c means the entrainment of the
system to the forcing frequency: frequency locking.
There are two effects on the frequency of system: forcing and nonlinear 
coupling\cite{comm}. Forcing, represented by 
$c\sqrt{1+b^2} \cos(2\phi + \delta)$, compensates the
term $(b-\nu)$ and entrains the system.
Thus, without the nonlinear coupling $-(b+d)(\nabla \phi)^2$, 
a frequency-locked stationary solution occurs for 
$c > |b-\nu|/\sqrt{1+b^2}$.
In other words, forcing increases the frequency for $b-\nu<0$ 
and decreases the frequency for $b-\nu>0$.
However, nonlinear coupling always increases the frequency of
the system by $-(b+d)k^2$ \cite{comm2}. 
Hence, for high $\nu$, forcing and nonlinear coupling effects 
are cooperating and frequency locking is enhanced. 
But for low $\nu$, two effects are competing and 
locking is suppressed.

These effects also explain the following transition routes to
locking as increasing the external forcing amplitude c.
(See Fig.\ref{fig1})

low $\nu$ : spiral $\rightarrow$ chaotically bursting domain 
		$\rightarrow$ frequency-locked front
		
high $\nu$ : spiral $\rightarrow$ labyrinth ($n=2$), target
		($n=3$) $\rightarrow$ frequency-locked front

At low $\nu$, the turbulent states, characterized by
chaotic nucleation and annihilation of domains, arise
between spirals and locked states(Fig.\ref{fig2}(f)).
As explained above, the diffusion effect against locking 
increases with $(\nabla \phi)^2$ at low $\nu$.
Near the defect core, $(\nabla \phi)^2$ is large and
frequency locking is suppressed.
However, far from the defect core, $(\nabla \phi)^2$ is 
small and domains are frequency-locked.
As the defects and the domains move, the dynamics become quite complex.
The chaotic time series of Re($\psi$) at one typical point, shown in 
Fig.\ref{fig4}(a), consists of flat parts and irregular
oscillating parts. 
Time evolution of $(\nabla \phi)^2$ at the same point is also
given in Fig.\ref{fig4}(b). 
One can observe that irregular oscillating parts correspond to 
large $(\nabla \phi)^2$.
Further increasing the external forcing, the domain pervades
the entire region and frequency-locked patterns form.

At high $\nu$, increasing the amplitude $c$ gives labyrinths
or targets. 
The labyrinths have a finite $(\nabla \phi)^2$,
which helps the frequency locking now. So, locked labyrinths
can arise outside the Arnol'd tongue of a single oscillator.
But the targets at 3:1 resonance are not frequency locked.
Although it might be due to the chirality, 
the origin of unlocked target patterns is not clear at this
stage and should be studied further. 

In summary, we examined frequency-locking phenomena 
in spatially extended systems using a complex 
Ginzburg-Landau equation with a resonant term.
A resonant tongue structure was constructed and 
studied as a function of $\nu$(measuring frequency mismatch) 
and $c$ (amplitude of forcing). Analysis with the phase equation 
showed that frequency locking is enhanced by diffusion at higher 
forcing frequencies, but suppressed at lower forcing frequencies.
This explains the locus of locked labyrinths
and unlocked chaotically bursting domains in parameter
space with respect to Arnol'd tongues.

The results are in agreement with recent experimental reports
\cite{pos97,lps99,lbm00}. 
For example, labyrinths formed at higher frequencies 
than $\pi$ fronts; $2\pi /3$ fronts at 3:1 resonance 
move due to the chirality.
In addition, novel patterns such as chaotically bursting domains 
and target patterns were also observed during the transition 
to locking. It would be interesting to know whether 
these are confirmed experimentally as well.

The author wishes to thank H.-T.Moon, S.-O.Jeong and T.-W.Ko for
many stimulating discussions. 
This work was supported in part by the project HPC-COSE from
the Ministry of Science and Technology and in part by BK21 
program of the Ministry of Education in Korea.

\pagebreak
\newpage


\begin{figure}
\caption{ Phase diagram : frequency deviation $\nu$ versus
critical forcing amplitude $c$. 
(a) $n=1$, (b) $n=2$, (c) $n=3$.
Solid lines are critical amplitude $c_{lo}$
for a single oscillator.
The shaded regions are obtained from the numerical
integration of Eq. (\ref{fcgl})
1F: flat, 2F:$\pi$ fronts, 3F:$2\pi/3$ fronts,
BD: chaotically bursting domains,
LABY: labyrinthine patterns, TA:targets.
See the text for details.}
\label{fig1}
\end{figure}

\begin{figure}
\caption{Numerical integration results, Plot of Re($\psi$),
(a) flat ($n=1, \nu=-0.4, c=0.09$).  
(b) $\pi$ fronts ($n=2, \nu=-0.1, c=0.4$). 
(c) $2 \pi /3$ fronts ($n=3, \nu=-0.1, c=0.45$).
(d) labyrinths ($n=2, \nu=0.1, c=0.4$). 
(e) target ($n=3, \nu=-0.15, c=0.32$). 
(f) chaotically bursting domains ($n=3, \nu=-0.65, c=0.26$). } 
\label{fig2}
\end{figure}

\begin{figure}
\caption{ (a) Time series at one point in Fig.\ref{fig2}(c).
(b) Parametric plot of real and imaginary parts of $\psi$ in a
complex plane corresponding to Fig.\ref{fig2}(c). The abscissa
is Re($\psi$) and the ordinate is Im($\psi$). }
\label{fig3}
\end{figure}

\begin{figure}
\caption{(a) Chaotic time series at one point in Fig.\ref{fig2}(f).
(b) Time evolution of $(\nabla \phi)^2$ at the same point.}
\label{fig4}
\end{figure}


\begin{references}
\bibitem{comm} 
Both the linear and nonlinear coupling term in Eq. (\ref{phase})
come from the diffusion term in Eq. (\ref{fcgl}).
Note that the linear term in Eq. (\ref{fcgl}) does not
contribute to the frequency significantly since the phase variable
$\phi$ takes a form like $\phi \sim \vec{k} \cdot \vec{r} $.
Hence only the nonlinear coupling term $-(b+d)(\nabla \phi)^2$  
reflects the diffusion effect on frequency in Eq. (\ref{fcgl}).
The interaction between phase and amplitude modes yields
the nonlinear coupling term from linear diffusive coupling.
In this sense, simple phase models such as coupled circle
map\cite{kaneko} cannot account for the results 
presented in this letter since it does not have amplitude variables.
\bibitem{comm2}
The term $-(b+d)(\nabla \phi)^2$ is always positive since $b+d<0$
from the experimental consideration explained near Eq. (\ref{phase0}). 
Note that labyrinths would arise at the lower frequency if we chose 
the sign of $b,d$ positive.
In such a case, diffusion decreases the frequency of patterns, 
enhancing locking at low frequencies.
\end{references}
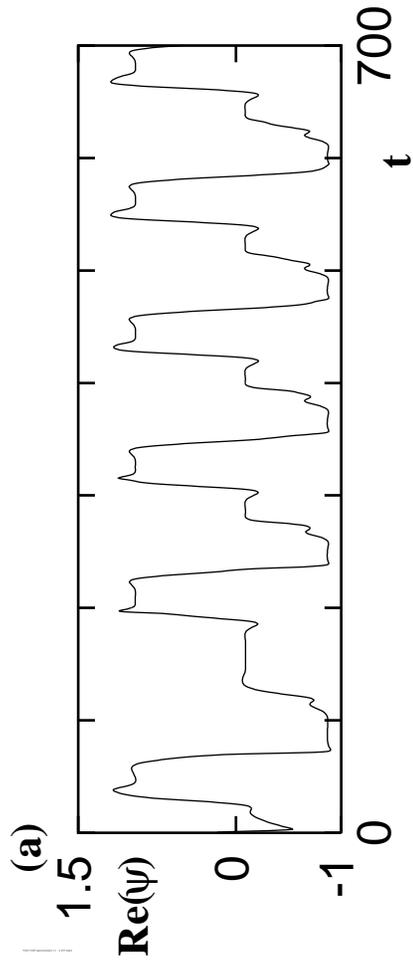
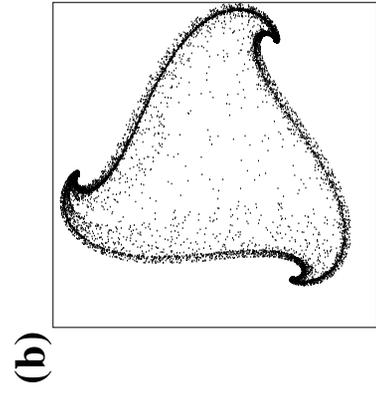
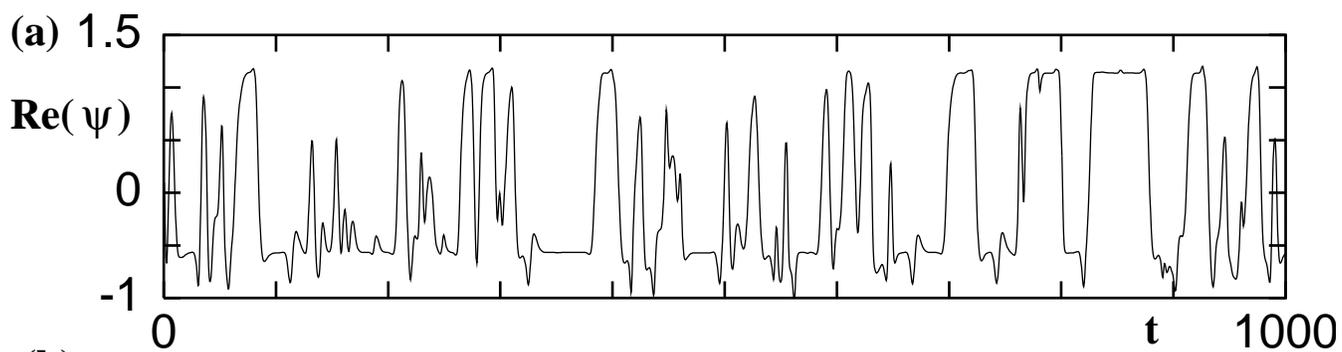
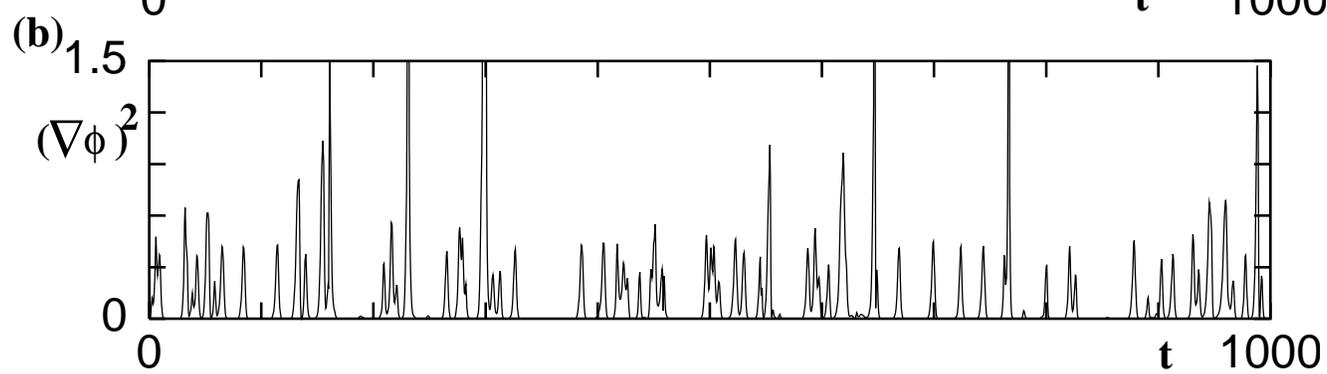
\end{document}